# ADAPTIVE MATCHING OF THE IOTA RING LINEAR OPTICS FOR SPACE CHARGE COMPENSATION*

A. Romanov†, A. Valishev, Fermilab, Batavia, USA
D. L. Bruhwiler, N. Cook, C. Hall, RadiaSoft LLC, Boulder, USA

*Abstract*

Many present and future accelerators must operate with high intensity beams when distortions induced by space charge forces are among major limiting factors. Betatron tune depression of above approximately 0.1 per cell leads to significant distortions of linear optics. Many aspects of machine operation depend on proper relations between lattice functions and phase advances, and can be improved with proper treatment of space charge effects. We implement an adaptive algorithm for linear lattice rematching with full account of space charge in the linear approximation for the case of Fermilab's IOTA ring. The method is based on a search for initial second moments that give closed solution and, at the same time, satisfy predefined set of goals for emittances, beta functions, dispersions and phase advances at and between points of interest. Iterative singular value decomposition based technique is used to search for optimum by varying wide array of model parameters.

## INTRODUCTION

The Integrable Optics Test Accelerator (IOTA) is under construction at Fermilab [1-3]. Figure 1 shows the IOTA ring with its main components. First stage experiments with electrons will allow to study in detail single particle dynamics in proposed nonlinear lattices. Second stage experiments with high intensity proton beams will demonstrate benefits of integrable insertions in case of operations with intense beams. Space charge forces alter the lattice structure and without proper adjustments can break the integrability of the accelerator. This work describes a rematching method that allows to find lattice configurations in which optimal conditions for nonlinear insertions are achieved for some predefined strength of space charge forces.

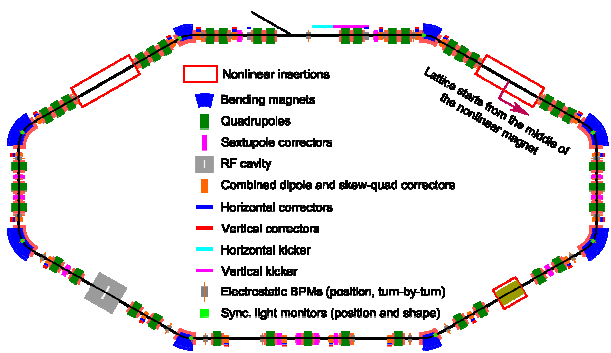

Figure 1: Schematic IOTA layout with its main components

___________________
*This work was partially supported by the US DOE Office of Science, Office of High Energy Physics, under Award Number DE-SC0011340.
† aromanov@fnal.gov

## MATCHING METHOD

Only linear effects are considered in this work, which is applicable either to specially prepared bunches with Kapchinskij–Vladimirskij distribution or to particles from the core of the beam. The goal is to find such lattice configuration that will have a closed solution for a given emittance and current in presence of space charge forces. At the same time, the lattice should satisfy requirements needed for integrability.

Strong space charge effects can make intrabeam dynamics unstable in a ring tuned for a beam with negligible space charge forces. Therefore, many traditional fitting software fail at the first step of calculation of initial closed solution. There are several workarounds, for example one can gradually increase beam current while fitting necessary parameters at each step, but such approaches require a lot of manual manipulations. To overcome initial stability issues, the algorithm was developed to treat the ring lattice as a channel with some initial guess on the beam's second moments at the beginning $M_{start}$. In order to have a closed solution, the set of goal parameters necessary for proper ring operation is expanded with requirement to have the same second moments at the beginning of the lattice and at the end:

$$M_{start} - T M_{start} T^t \to 0, \quad (1)$$

where $T$ is a transport matrix of a full lattice with self-forces accounted.

In case of arbitrarily coupled beam, full matrix of second moments must be matched to have a closed solution (1). For the coasting beam there is no constraints on the time correlations and, therefore, elements in the fifth row and the fifth column can differ:

$$(M_{start} - T M_{start} T^t) \cdot \mathrm{Diag}[1,1,1,1,0,1] \to 0, \quad (2)$$

For a coasting beam in a ring without transverse coupling, the proposed approach adds 6 free and 6 goal parameters: beta functions ($\beta_x$, $\beta_y$), horizontal dispersion ($D_x$) and its primes at the beginning and the end of the treated lattice.

Experiment with one nonlinear magnet in IOTA ring has a set of requirements to the linear lattice, including exact values of lattice functions in the middle of the nonlinear insertion. Moving the beginning of the lattice to the middle of special magnet, automatically reduce number of goals and number of free parameters.

The beam-based lattice fitting algorithm that was developed and implemented in *sixdsimulation* software [4] was adapted to solve the described problem and demonstrated ability to find solutions for self-forces significantly exceeding stability region of a lattice tuned for zero current. In order to account for the linear part of space charge forces, a matrix of second moments is traced in small steps with lengths much smaller than beta functions in both planes. At each step, an additional transport matrix corresponding to a thin defocusing kick is applied with a strength equivalent to the integrated effect from self-forces.

Space charge forces only affects intrabeam dynamics, while beam trajectory and dynamics of tail particles are determined only by external fields. Therefore, even for short term stability, it is crucial to have stable closed solutions for both cases of full and zero space charge forces.

There are several ways to define initial second moments. The most straightforward approach is direct enumeration of all 21 elements of symmetric 6x6 matrix. In order to calculate Twiss parameters along the ring, the eigen-system of matrix *SM* must be calculated, where S is simplectic matrix. Eigenvalues multiplied by complex *i* will give emittances and other lattice parameters can be derived from eigenvectors.

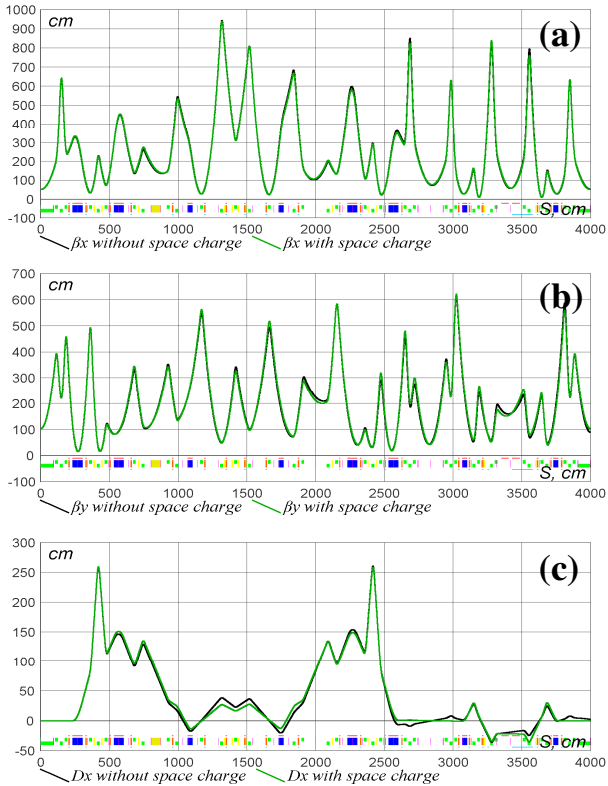

Figure 2: Horizontal beta functions (a), vertical beta functions (b) and horizontal dispersions (c) with (green) and without (black) space charge forces for experiment with one nonlinear magnet set to strength *t=0.3* in IOTA ring treated as a channel; space charge tune depression is -0.03

In most cases, it is known from the symmetries of the design that some second moments of the beam must be zero and some have internal correlations. In such situations, it might be reasonable to define initial betas, dispersions, and its primes along with the desired emittances. This simplified approach allows to simultaneously construct a matrix of the beam's second moments and mode vectors to calculate Twiss parameters along the ring. For a found closed lattice configuration, eigenvectors will coincide with mode vectors.

*Rematching in presence of nonlinear insertion*

Nonlinear insertions planned for IOTA, such as special iron-based magnet or electron lens, will have fields with linear focusing components. For the case of zero current approximation, these forces may be omitted in linear lattice models because of the way the requirements to such lattices are set. For a nonlinear experiment, IOTA ring should form some specific linear transformation from one end of the insertion to another. The required transformation can be described either as a set of Twiss parameters or as a specific transport matrix of the exit-to-enter arc. In case of non-zero space charge, linear forces of non-linear insertion will change beam envelope and corresponding space charge forces in the whole ring. Therefore, matching must be done for every used setup of special insertions. In other words, in a presence of self-forces, a non-linear insertion changes transport matrix in the outer arc and thus can't be omitted anymore.

## RESULTS

All calculations of re-matched lattices presented here were done for the IOTA lattice tuned for experiment with one nonlinear magnet. The beam was assumed to be coasting and, therefore, the set of requirements to the initial and final second moments is determined by (2). The beginning of the lattice is set at the middle of the nonlinear magnet (see Figure 1) and initial conditions are fixed to the values extracted from the lattice tuned for zero space charge forces, but with the account for quadrupole component of nonlinear element set to the desired strength of *t=0.3*.

Because of initial conditions fixed to correct values, there are 20 variable parameters and 8 constraints. The variables are the gradients in quadrupoles and the constraints are the 2 phase advances from exit to entrance of the nonlinear magnet plus the 6 parameters needed for a closed solution ($\beta_x$, $\alpha_x$, $\beta_y$, $\alpha_y$, $D_x$, $D'_x$).

For the initial simulations with *Synergia* [5], it was decided to focus on a task with isolated problem of space charge compensation. To have space charge forces as small distortion and avoid chromatic effects, tunes depressions were selected to be close to -0.03 and momentum spread was set to zero. Corresponding rematched lattice has small deviations from lattice tuned for zero current approximation. But even for such a small correction, lattice functions for zero current beam in a rematched ring differ from optimal by more than maximal errors allowed for integrable lattices. Results are presented in Figure 2 and Table 1.

*Synergia* analysis consisted of running a single turn with a matched distribution with explicitly linear space charge solver turned on. The bunch has a generalized KV distribution. Particle coordinates are output at the exit of the nonlinear insert and again at the entrance. A linear phase unwrap algorithm is then applied to the particles normalized coordinates at those positions and a mean phase advance (modulo 2π) is computed for that segment of the lattice. Figure 3 shows current dependence of reconstructed phase advances in transverse planes for the normalized KV-distribution with $H_0 = 4$ mm-mrad.

To illustrate capabilities of the developed method, a rematching was done for the case of space charge tune depression around -0.45. Figure 4 shows plots of beta functions and dispersions with and without space charge in a channel mode.

Table 1: Errors of lattice parameters in the case of beam with zero current in lattice rematched for tunes depressions of about -0.03

|  | $\beta_x$, % | $\beta_y$, % | $D_x$, cm | $\nu_x$ | $\nu_x$ |
|---|---|---|---|---|---|
| **Max err** | 1.0 % | 1.0 % | 1.0 | 0.001 | 0.001 |
| **Rematch err** | -3.2 % | -1.8 % | 1.4 | 0.033 | 0.030 |

## CONCLUSION

Even though realistic tracking over one turn confirms compensation of linear space charge forces, it is necessary to address the found mismatch between horizontal and vertical planes and also directly study multi-turn stability to prove benefits from the proposed lattice modifications. Furthermore, the complexity of the task requires incremental introduction of features that might affect beam dynamics. For example, energy spread, frozen or dynamically updated space charge forces, longitudinal dynamics, etc. At each step, it is important to cross-check results with several simulation tools.

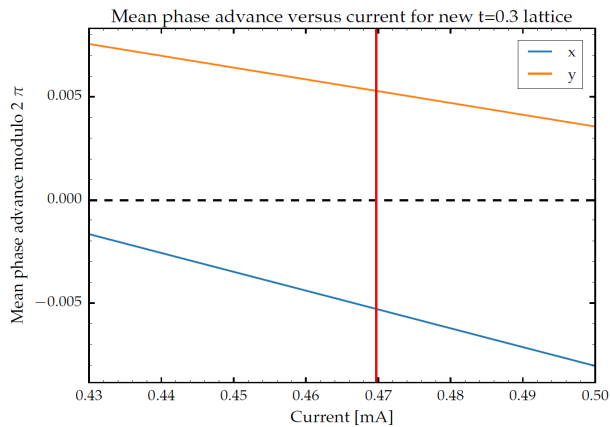

Figure 3: Mean phase advance from exit of the nonlinear insertion to its entrance, obtained from *Synergia* simulations

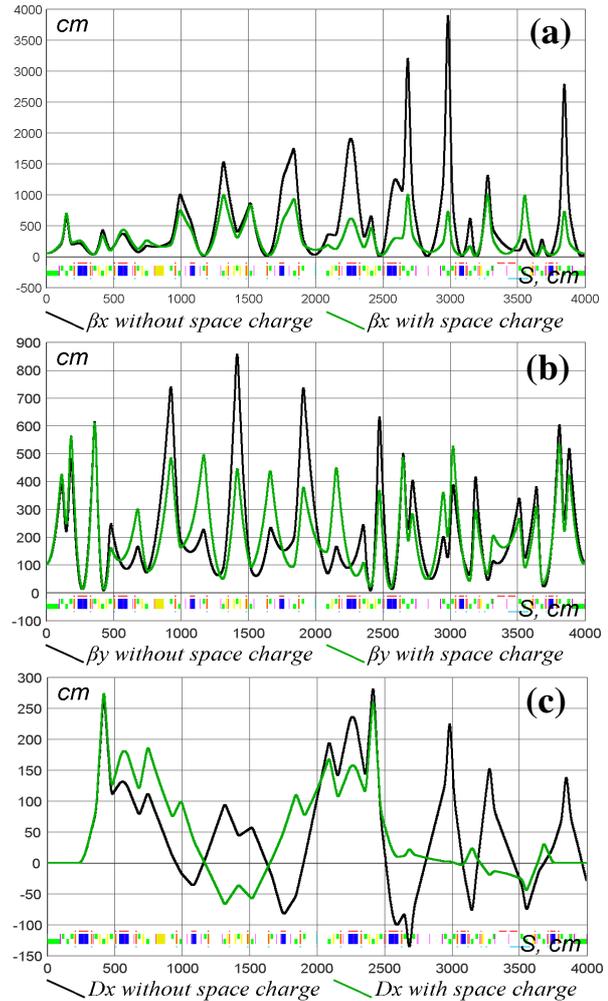

Figure 4: Horizontal beta functions (a), vertical beta functions (b) and horizontal dispersions (c) with (green) and without (black) space charge forces for experiment with one nonlinear magnet set to strength t=0.3 in IOTA ring treated as a channel; space charge tune depression -0.45